\documentclass[twocolumn,prb]{revtex4}
\usepackage{amssymb}
\usepackage{amsmath}
\usepackage{graphicx}

\input{tcilatex}

\begin{document}

\title{Resonant spin Hall conductance in quantum Hall systems lacking bulk
and structural inversion symmetry}
\author{Shun-Qing Shen$^{1}$, Yun-Juan Bao$^{1}$, Michael Ma$^{2}$, X.C. Xie$
^{3,4}$, and Fu Chun Zhang$^{1,2,5}$}
\affiliation{$^{1}$Department of Physics, The University of Hong Kong, Pukfulam Road,
Hong Kong, China\\
$^{2}$Department of Physics, University of Cincinnati, Ohio 45221\\
$^{3}$Department of Physics, Oklahoma State University, Stillwater, Oklahoma
74078\\
$^{4}$ICQS, Institute of Physics, Chinese Academy of Sciences, Beijing,
China \\
$^{5}$Department of Physics, Zhejiang University, Hangzhou, Zhejiang, China }
\date{October 7, 2004}

\begin{abstract}
Following a previous work [ Shen, Ma, Xie and Zhang, Phys. Rev. Lett. 92,
256603 (2004)] on the resonant spin Hall effect, we present detailed
calculations of the spin Hall conductance in two-dimensional quantum wells
in a strong perpendicular magnetic field. The Rashba coupling, generated by
spin-orbit interaction in wells lacking bulk inversion symmetry, introduces
a degeneracy of Zeeman-split Landau levels at certain magnetic fields. This
degeneracy, if occuring at the Fermi energy, will induce a resonance in the
spin Hall conductance below a characteristic temperature of order of the
Zeeman energy. At very low temperatures, the spin Hall current is highly
non-ohmic. The Dresselhaus coupling due to the lack of structure inversion
symmetry partially or completely suppresses the spin Hall resonance. The
condition for the resonant spin Hall conductance in the presence of both
Rashba and Dresselhaus couplings is derived using a perturbation method. In
the presence of disorder, we argue that the resonant spin Hall conductance
occurs when the two Zeeman split extended states near the Fermi level
becomes degenerate due to the Rashba coupling and that the the quantized
charge Hall conductance changes by $2e^2/h$ instead of $e^2/h$ as the
magnetic field changes through the resonant field.
\end{abstract}

\maketitle

\section{Introduction}

Spintronics, which exploits electron spin rather than charge to develop a
new generation of electronic devices, has emerged as an active field in
condensed matters because of both the underlying fundamental physics and its
potential impact on the information industry.\cite%
{Prinz98Science,Wolf01Science,Awschalom02} One key issue in spintronics is
the generation and efficient control of spin current. Spin-orbit interaction
of electrons exists extensively in metals and semiconductors and mix spin
states. It provides an efficient way to control the coherent motion of
electron spins. Recently it is proposed theoretically that an electric field
may generate a spin current in hole-doped semiconductors and in
two-dimensional electron gases (2DEG) in heterostructures with spin-orbit
coupling due to the spin helicity and the noncollinearity of the velocity of
the single particle wave function.\cite{Murakami03Science,Sinova04,Shen04prl}
Studies of this intrinsic spin Hall effect has evolved into a subject of
intense research. \cite%
{Shen03xxx,Sinitsyn03xxx,Culcer03xxx,Schliemann04prb,Hu03xxx,Rashba03prb,Hulb04xxx}
The spin Hall effect in a paramagnetic metal with magnetic impurities has
also been discussed, in which a transverse spin imbalance will be generated
when a charge current circulates.\cite%
{Dyakonov71,Hirsch99,Zhang00prl,Hu03prb} We also note that the spin
chirality in systems with strong spin-orbit interaction may induce a pure
spin current\cite{Shen97pla}

Over the past two decades, remarkable phenomena have been observed in the
2DEG, most notably, the discovery of integer and fractional quantum Hall
effect.\cite{Klitzing80,Tsui82,Prange87} Research in spin transports
provides a good opportunity to explore spin physics in the 2DEG with
spin-orbit couplings. The spin-orbit coupling leads to a zero-field spin
splitting, and it competes with the Zeeman spin splitting when a
perpendicular magnetic field is applied. The result can be detected as
beating in Shubnikov-de Haas oscillations.\cite{Nitta97prl,Heida98prb}

Very recently we have studied the spin Hall effect in the 2DEG with
spin-orbit coupling in a strong perpendicular magnetic field, and predicted
a resonant spin Hall effect caused by the Landau level crossing near the
Fermi energy.\cite{Shen04prl} In this paper we present detailed calculations
of the problem. We analyze symmetries in systems with the Rashba and/or
Dresselhaus couplings. By using linear response theory, we calculate the
spin Hall conductance $G_{s}$, including its magnetic field and temperature
dependences for realistic parameters of InGaGs/InGaAlGs. The non-linearity
in the electric field of the spin Hall current near resonance is also
studied beyond the linear response theory. The resonance is a low
temperature property, which shows up at a characteristic temperature of the
order of the Zeeman energy $E_{Z}$. The peak of the resonance diverges as $%
1/\max (k_{B}T,eEl_{b})$ ($l_{b}$: the magnetic length), and its weight
diverges as $-\ln T$ at low $T$ and at $E\rightarrow 0$. Near the resonant
magnetic field $B_{0}$, $G_{s}\propto 1/\left\vert B-B_{0}\right\vert $. The
resonance arises from the Fermi level degeneracy of the Zeeman-split Landau
levels in the presence of the spin-orbit coupling. Among the two types of
the spin couplings we consider, the Rashba coupling reduces the Zeeman
splitting and is the interaction responsible for the resonance. The
Dresselhaus coupling further separates the Zeeman splitting and suppresses
the resonance. The resonant condition in the presence of both Rashba and
Dresselhaus couplings is derived within a perturbation theory, which is
accurate for small ratio of the Zeeman energy to the cyclotron frequency.

The paper is organized as follows. In Section II we introduce the
Hamiltonian of the system under consideration and analyze its symmetries. In
Section III, we study the spin Hall current for systems with only Rashba or
only Dresselhaus coupling. In Section IV, we consider systems with both
Rashba and Dresselhaus couplings. By treating the couplings as small
parameters, we develop a perturbation method to derive the resonance
condition. The paper is concluded with a summary and discussions in Section
V.

\section{Model Hamiltonian and symmetry}

\subsection{Spin orbit coupling and model Hamiltonian}

As an introduction, we start with the three-dimensional (3D) spin-orbit
interaction known for III-V compounds such as GaAs and InAs, which is of the
form\cite{Dresselhaus55,Rashba60} 
\begin{equation}
V_{so}^{3D}=\alpha _{0}\mathbf{K}(\mathbf{p})\cdot \mathbf{\sigma }+\beta
_{0}\mathcal{E}\cdot \mathbf{(p}\times \mathbf{\sigma })
\end{equation}%
where $\sigma _{\mu }$ ($\mu =x,y,z$) are the Pauli matrices for spin of
electrons, $\mathbf{p}$ is the momentum of the charge carrier, and 
\begin{equation}
K_{\mu }(\mathbf{p})=\sum_{\nu ,\delta }p_{\nu }p_{\mu }p_{\nu }\epsilon
_{\mu ,\nu ,\delta }.
\end{equation}%
In Eq.(1), the first term is the Dresselhaus coupling which originates from
the lack of bulk inversion symmetry,\cite{Dresselhaus55} while the second
term is the Rashba coupling which arises from the lack of structure
inversion symmetry.\cite{Rashba60} The effective field $\mathcal{E}$ is
induced by the asymmetry of the external voltage to the system. In quantum
wells, by neglecting the weak interband mixing and retaining the linear
contribution of $\mathbf{p}$ parallel to the $x-y$ plane, the spin-orbit
interaction in 3D is reduced to an effective one in 2D, 
\begin{subequations}
\begin{eqnarray}
V_{so}^{2D} &=&H_{so}^{D}+H_{so}^{R} \\
H_{so}^{D}(\alpha ) &=&\frac{\alpha }{\hbar }(\sigma _{x}p_{x}-\sigma
_{y}p_{y})  \label{Dresselhaus} \\
H_{so}^{R}(\beta ) &=&\frac{\beta }{\hbar }(\sigma _{y}p_{x}-\sigma
_{x}p_{y})  \label{Rashba}
\end{eqnarray}%
where $\alpha =-\alpha _{0}\hbar \left\langle p_{z}^{2}\right\rangle $ and $%
\beta =\beta _{0}\hbar \left\langle \mathcal{E}_{z}\right\rangle $, with the
average taken over the lowest energy band of the quasi-2D quantum well. The
Rashba coupling can be modulated up to fifty percent by a gate voltage
perpendicular to the plane\thinspace \cite{Nitta97prl,Grundler00}. In some
quantum wells such as GaAs the two terms are usually of the same order of
magnitude, while in narrow gap compounds like InAs the Rashba coupling
dominates.\cite{Jusserand95,Knap96,Miller03} Experimentally the relative
strength of the Rashba and Dresselhaus couplings can be extracted from
photocurrent measurements. \cite{Ganichev04prl}

\begin{figure}[tbp]
\includegraphics[width=8.5cm]{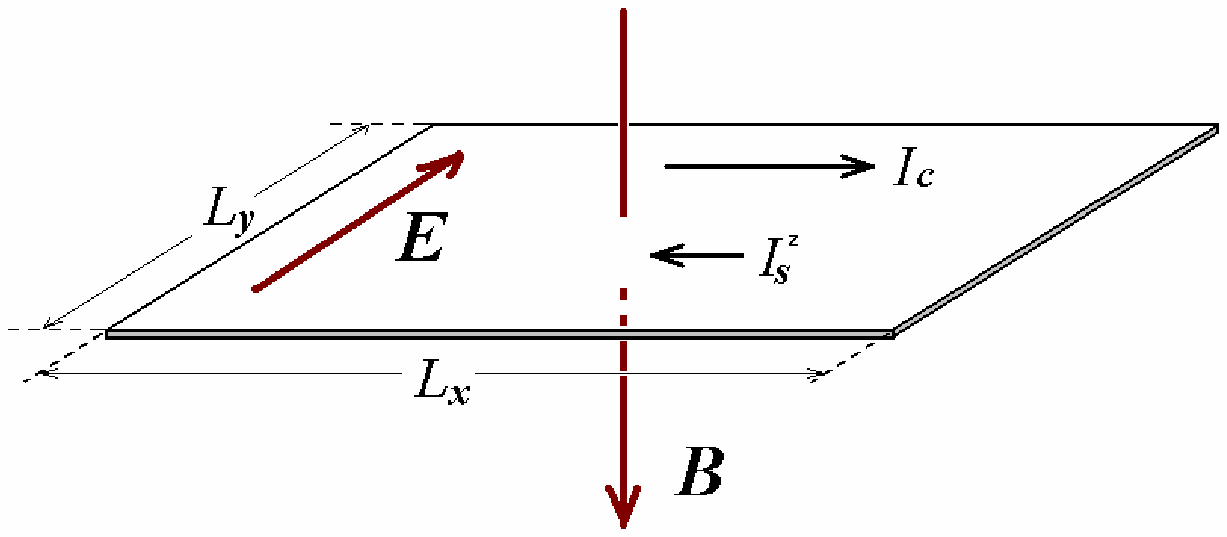}
\caption{{Schematic illustration of the two-dimensional electron gas studied
in the text. $I_{e}$ and $I_{s}$ are the charge and spin Hall currents.}}
\end{figure}

In this paper we consider a spin-1/2 particle of charge $-e$ and effective
mass $m$ confined by a semiconductor quantum well to a 2D $x-y$ plane of
length $L_{x}$ and width $L_{y}$. \cite{note1} The particle is subjected to
a spin-orbit interaction $V_{so}^{2D}$. A perpendicular magnetic field $%
\mathbf{B}=-B\hat{z}=\nabla \times \mathbf{A}$ and an electric field $%
\mathbf{E}=E\hat{y}$ along the y-axis are applied as shown in Fig. 1. Both
electron-electron interaction and impurities will be neglected in our study.
The Hamiltonian reads 
\end{subequations}
\begin{eqnarray}
H &=&H_{0}+eEy,  \notag \\
H_{0} &=&\frac{1}{2m}\left( \mathbf{p}+\frac{e}{c}\mathbf{A}\right) ^{2}- 
\frac{1}{2}g_{s}\mu _{B}B\sigma _{z}+V_{so}^{2D}(\mathbf{A})  \label{H}
\end{eqnarray}
where $g_{s}$ is the Lande g-factor, and $\mu _{B}$ is the Bohr magneton. In 
$V_{so}^{2D}(\mathbf{A})$ the momentum $\mathbf{p}$ is replaced by the
canonical momentum, $\Pi =\mathbf{p}+\frac{e}{c}\mathbf{A}$. We choose the
Landau gauge $\mathbf{A}=yB\hat{x}$ and consider periodic boundary condition
in the $x$ direction, hence $p_{x}=k$ is a good quantum number.

Below we rewrite the Hamiltonian in terms of lowering and raising operators.
For each $k$, we introduce the lowering operator 
\begin{equation*}
a_{k}=\frac{1}{\sqrt{2}l_{b}}\left[ y+\frac{c}{eB}(k+ip_{y})\right]
\end{equation*}%
and the corresponding raising operator $a_{k}^{\dag }=(a_{k})^{\dag }$, with
the magnetic length $l_{b}=\sqrt{\hbar c/eB}$. $a$ and $a^{\dag }$ satisfy
the commutations $\left[ a_{k},a_{k^{\prime }}^{+}\right] =\delta
_{kk^{\prime }}$, and $\left[ a_{k},a_{k^{\prime }}\right] =0$. In terms of $%
a_{k}$ and $a_{k}^{+},$ we have 
\begin{eqnarray}
H_{0}/\hbar \omega &=&a_{k}^{+}a_{k}+\frac{1}{2}(1-g\sigma _{z})+i\sqrt{2}%
\eta _{R}\left( a_{k}\sigma _{-}-a_{k}^{+}\sigma _{+}\right)  \notag \\
&&+\sqrt{2}\eta _{D}\left( a_{k}^{+}\sigma _{-}+a_{k}\sigma _{+}\right)
\label{hh}
\end{eqnarray}%
where $\omega =eB/mc$ is the cyclotron frequency, $\sigma _{\pm }=(\sigma
_{x}\pm i\sigma _{y})/2$, and $g=g_{s}m/2m_{e}$ is twice of the ratio of the
Zeeman energy to the cyclotron frequency ($m_{e}$: free electron mass). $%
\eta _{R}=\beta ml_{b}/\hbar ^{2}$ and $\eta _{D}=\alpha ml_{b}/\hbar ^{2}$,
both inversely proportional to $\sqrt{B}$ are the dimensionless Rashba and
Dresselhaus coupling respectively.

The velocity operator plays an important role in study of transport
properties including the spin Hall conductance. The velocity operator of a
single particle is $v_{\tau }=[\tau ,H]/i\hbar $ ($\tau =x,y$), from which
we obtain 
\begin{subequations}
\begin{eqnarray}
v_{x} &=&\frac{\hbar }{\sqrt{2}ml_{b}}\left[ a_{k}^{+}+a_{k}+\sqrt{2}\eta
_{D}\sigma _{x}+\sqrt{2}\eta _{R}\sigma _{y}\right] , \\
v_{y} &=&\frac{i\hbar }{\sqrt{2}ml_{b}}\left[ a_{k}^{+}-a_{k}+i\sqrt{2}\eta
_{D}\sigma _{y}+i\sqrt{2}\eta _{R}\sigma _{x}\right] .
\end{eqnarray}
Comparing this with the standard expression of velocity for a charged
particle in a magnetic field, $\mathbf{v}=(\mathbf{p}+\frac{e}{c}\mathbf{A}%
)/m,$ the spin-orbit coupling effectively induces a spin-dependent vector
potential.

\subsection{Symmetries}

We analyze three symmetries of the Hamiltonian in this subsection, which we
will use in our calculations.

\textit{Interchange symmetry of the two couplings. }Under the unitary
transformation, $\sigma _{x}\rightarrow \sigma _{y},$ $\sigma
_{y}\rightarrow \sigma _{x},$ $\sigma _{z}\rightarrow -\sigma _{z},$ the
Rashba and Dresselhaus couplings are interchanged\cite{Shen03xxx}, 
\end{subequations}
\begin{subequations}
\begin{eqnarray}
\alpha (\Pi _{x}\sigma _{x}-\Pi _{y}\sigma _{y}) &\rightarrow &\alpha (\Pi
_{x}\sigma _{y}-\Pi _{y}\sigma _{x}); \\
\beta (\Pi _{x}\sigma _{y}-\Pi _{y}\sigma _{x}) &\rightarrow &\beta (\Pi
_{x}\sigma _{x}-\Pi _{y}\sigma _{y}); \\
g_{s} &\rightarrow &-g_{s}.
\end{eqnarray}%
Therefore a system with Rashba coupling $\beta $, Dresselhaus coupling $%
\alpha $, and Lande g-factor $g_{s}$ is mapped on to a system with Rashba
coupling $\beta $, Dresselhaus coupling $\alpha $, and Lande g-factor $%
-g_{s} $. In particular, a system with only Dresselhaus coupling can be
mapped on to a system with only Rashba coupling and an opposite sign in $%
g_{s}$. This symmetry will be used in Section III. At the symmetric point $%
\alpha =\beta $, $V_{so}^{2D}$ is invariant under the transformation. $%
\alpha =-\beta $ is another symmetric point under the transformation, $%
\sigma _{x}\rightarrow -\sigma _{y},$ $\sigma _{y}\rightarrow -\sigma _{x},$ 
$\sigma _{z}\rightarrow -\sigma _{z}.$ For physical parameters, we will
always consider $g_{s}>0$.

\textit{Signs of the couplings. }Under the transformation, $\sigma
_{x}\rightarrow -\sigma _{x},$ $\sigma _{y}\rightarrow -\sigma _{y},$ $%
\sigma _{z}\rightarrow \sigma _{z},$we have $\alpha \rightarrow -\alpha $
and $\beta \rightarrow -\beta $. The eigenenergy spectrum is invariant under
the simultaneous sign changes of the two couplings. The eigenenergy spectrum
is even in $\eta _{R}$ if $\eta _{D}=0$ and is even in $\eta _{D}$ if $\eta
_{R}=0$.

\textit{Charge conjugation.} Under the charge conjugation transformation, $
-e\rightarrow e$, the magnetic moment of the carrier also changes its sign,
or effectively $g_{s}\rightarrow -g_{s}$ in Eq. (4). This transformation is
equivalent to the flip of the external magnetic field $B\rightarrow -B$.
Therefore, a system of hole carriers has the same physical properties as the
corresponding electron system except for possible directional changes in the
observables.

$H_{0}$ can be solved analytically in the systems with only Rashba or only
Dresselhaus coupling. An analytical solution is currently not available for $
H_{0}$ with both couplings. \cite{Das90,Falko93,Schliemann03prb} In the next
section, we shall discuss the charge and spin Hall conductance of the
electron system with a pure Rashba coupling. The results can be mapped
easily onto the system with a pure Dresselhaus coupling and to the hole
system in semiconductors by using the symmetries discussed above.

\section{Systems with pure Rashba coupling}

In this section we focus on systems with the Rashba coupling only. After a
brief review of the single particle solution in the absence of an electric
field, we will discuss the spin Hall conductance by using linear response
theory in IIIB, and its non-linear effect and scaling behavior near the
resonance in IIIC. We will use the interchange symmetry to comment on the
Dresshaus coupling system.

\subsection{Single particle solution}

The single particle problem of $H_{0}$ with $\eta _{D}=0$ can be solved~\cite%
{Rashba60}. The Rashba coupling hybridizes a spin down state in the $
n_{0}^{th}$ Landau level with a spin-up state in the $(n_{0}+1)^{th}$ Landau
level, and the eigenenergies are given by 
\end{subequations}
\begin{equation}
\epsilon _{ns}^{R}=\hbar \omega \left( n+\frac{s}{2}\sqrt{(1-g)^{2}+8n\eta
_{R}^{2}}\right)  \label{r-spectrum}
\end{equation}
with $s=\pm 1$ for positive integer $n$, and $\epsilon _{0,+}=$ $\hbar
\omega (1-g)/2$. There is a large degeneracy $N_{\phi }=L_{x}L_{y}/(2\pi
l_{b}^{2})$ to each eigenergy. The corresponding eigenstates are given by 
\begin{equation}
\left\vert n,k,s\right\rangle =\left( 
\begin{array}{c}
\cos \theta _{ns}\phi _{nk} \\ 
i\sin \theta _{ns}\phi _{n-1k}%
\end{array}
\right)
\end{equation}
where $\phi _{nk}$ is the eigenstate of the n$^{th}$ Landau level with $
p_{x}=k$ in the absence of the spin-orbit coupling. $\theta _{0+}=0,$ and $
\tan \theta _{ns}=-u_{n}+s\sqrt{1+u_{n}^{2}}$ for $n\geq 1$, with $
u_{n}=(1-g)/\sqrt{8n\eta _{R}^{2}}$.

The eigenenergies for the system with Dresselhaus coupling only can be
obtained by replacing $\eta _{R}$ by $\eta _{D}$ and $g$ by $-g$, 
\begin{equation}
\epsilon _{ns}^{D}=\hbar \omega \left( n+\frac{s}{2}\sqrt{(1+g)^{2}+8n\eta
_{D}^{2}}\right) .  \label{d-spectrum}
\end{equation}
The energy spectra versus $\eta _{R}$ or $\eta _{D}$ are plotted in Fig. 2.
In the absence of the spin-orbit coupling, the Zeeman energy splits the two
degenerate $n_{0}^{th}$ Landau levels of spin-up and spin-down electron
states into two nearby ones with the lower level for spin-up and the higher
level for spin-down. As $\eta _{R}$ increases from zero, the energy of the $
n_{0}^{th}$ Landau level state of spin-down is lowered because of its
hybridization with the spin-up state at the $(n_{0}+1)^{th}$ Landau level
due to the Rashba coupling. The Rashba interaction competes with the Zeeman
energy and there is an energy crossing at certain values of $\eta _{R}$ or
the magnetic fields as we can see in Fig. 2(a). The spin Hall resonance we
examine is closely related to this level crossing. The energy level diagram
in Fig. 2(b) for the Dresselhaus coupling has different features. In that
case, a spin-up state, which is at the lower level due to the Zeeman
splitting, mixes with a spin-down state at a higher Landau level, which
separate further the Zeeman splitting, thus there is no resonance in the
spin Hall current.

\begin{figure}[tbp]
\includegraphics[width=8.5cm]{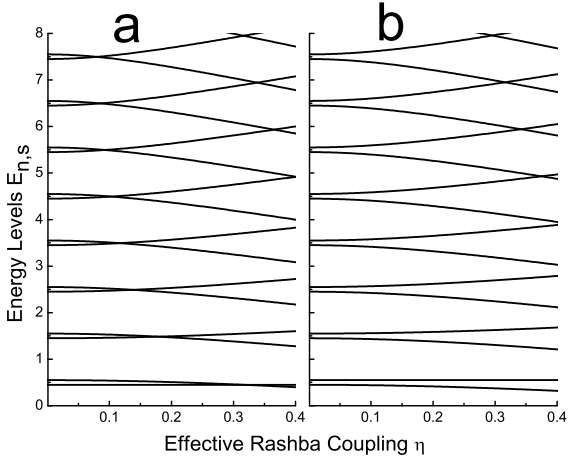}
\caption{{(a). Energy levels in unit of $\hbar \protect\omega $ as a
function of the dimensionless Rashba coupling $\protect\eta _{R}$. The
parameters are $\protect\beta =0.9\times 10^{-11}$ eVm, $n_{e}=1.9\times
10^{16}/$m$^{2}$, $m=0.05m_{e}$ and $g_{s}=4$, taken from Reference 
\protect\cite{Nitta97prl} for the inversion heterostructures In $_{0.53}$Ga$
_{0.47}$As/In$_{0.52}$Al$_{0.48}$As. (b) same as in (a), but for the
Dresselhaus coupling $\protect\eta_D$.} }
\end{figure}

\subsection{Linear response theory: Spin Hall conductance}

We consider the charge and spin Hall currents along the $x-$axis induced by
an electric field along the $y-$axis as in Fig. 1. In terms of the velocity
operator, the charge and spin-$z$ component current operators are defined by 
\begin{eqnarray}
j_{c} &=&-ev_{x}, \\
j_{s} &=&\frac{\hbar }{4}\left( \sigma _{z}v_{x}+v_{x}\sigma _{z}\right) ,
\end{eqnarray}%
respectively. We refer readers to Ref. ~\cite{Shen04prl} for the discussions
on the other spin components. The symmeterized form of the spin current
operator guarantees that it is Hermitian. Each single particle state $%
\left\vert \phi _{nks}\right\rangle $ carries a current $\left\langle \phi
_{nks}\right\vert j_{c,s}\left\vert \phi _{nks}\right\rangle $. The average
current density carried by the 2DEG is then given by 
\begin{equation}
I_{c,s}=\frac{1}{L_{x}L_{y}}\sum_{nks}f_{nks}\left\langle \phi
_{nks}\right\vert j_{c,s}\left\vert \phi _{nks}\right\rangle
\end{equation}%
where $f_{nks}$ is the Fermi-Dirac distribution function. Note that since
spin is not a conserved quantity in the presence of spin-orbit couplings,
the spin current defined above and the spin density do not satisfy a
continuity equation. Nevertheless, the expectation values of the spin
density and the spin current are well defined. The charge and spin Hall
conductance are then given by 
\begin{eqnarray}
G_{c} &=&I_{y}^{0}/E; \\
G_{s} &=&I_{y}^{z}/E.
\end{eqnarray}%
Unlike a free electron in an uniform magnetic field, the single particle
problem with the spin-orbit coupling in the presence of an electric field $E%
\hat{y}$ is not analytically solvable, since the Landau levels mixing no
longer truncates. After a replacement of $y\rightarrow y+eE/m\omega ^{2}$ in
the operator $a_{k}$ by $\tilde{a}_{k}=a_{k}+\frac{eEl_{b}}{\sqrt{2}\hbar
\omega },$ the Hamiltonian of the system in the presence of the electric
field reads, apart from a constant, 
\begin{equation}
H=H_{0}(E)+H^{\prime }
\end{equation}%
where $H_{0}(E)$ is the one in Eq. (\ref{hh}) by replacing $a_{k}$ by $%
\tilde{a}_{k}$ and $H^{\prime }=-eEl_{b}\eta _{R}\sigma _{y}$. We now
consider $H^{\prime }$ as a perturbative Hamiltonian to study the charge and
spin Hall currents. Up to the first order in $E$, we obtain 
\begin{equation}
(j_{c,s})_{nks}=(j_{c,s}^{(0)})_{nks}+(j_{c,s}^{(1)})_{nks}
\end{equation}%
where the superscript refers to the 0$^{th}$ order and 1$^{st}$ order in the
perturbation in $H^{\prime }$, and 
\begin{eqnarray*}
(j_{c,s}^{(0)})_{nks} &=&\left\langle n,k,s\right\vert j_{c,s}\left\vert
n,k,s\right\rangle ; \\
(j_{c,s}^{(1)})_{nks} &=&\sum_{n^{\prime }s^{\prime }}\frac{\left\langle
n,k,s\right\vert j_{c,s}\left\vert n^{\prime },k^{\prime },s^{\prime
}\right\rangle \left\langle n^{\prime },k^{\prime },s^{\prime }\right\vert
j_{c,s}\left\vert n,k,s\right\rangle }{\epsilon _{ns}^{R}-\epsilon
_{n^{\prime }k^{\prime }}^{R}} \\
&&+h.c.
\end{eqnarray*}%
In the above equations, the summation is over the complete and orthogonal
set of eigenstates $\{\left\vert n,k,s\right\rangle \}$ corresponding to the
spectra $\{\epsilon _{ns}^{R}\}$ in Eq. (8), and $n^{\prime }=n\pm 1$ since
the matrix elements vanish for other values of $n^{\prime }$. The charge
Hall conductance is found to be independent of the spin-orbit coupling, $%
G_{c}=\nu e^{2}/h$, with $\nu =N_{e}/N_{\phi }$ being the filling factor.
Within the perturbation theory, the spin Hall conductance $G_{s}$ can be
divided into two parts. The part arising from the $0^{th}$ order in $%
H^{\prime }$ is found to be the product of the spin polarization $%
\left\langle S^{z}\right\rangle $ per electron and the Hall conductance $%
G_{c}$, divided by the electron charge ($-e$), 
\begin{equation}
G_{s}^{(0)}=-\left\langle S^{z}\right\rangle G_{c}/e.
\end{equation}%
The expectation value of the spin polarization per electron is, 
\begin{eqnarray}
\left\langle S^{z}\right\rangle &=&\frac{1}{N_{e}}\frac{\hbar }{2}%
\sum_{nks}\left\langle n,k,s\right\vert \sigma _{z}\left\vert
n,k,s\right\rangle f_{nks}  \notag \\
&=&\frac{1}{N_{e}}\frac{\hbar }{2}\sum_{nks}\cos 2\theta _{ns}f_{nks}.
\end{eqnarray}%
$\left\langle S^{z}\right\rangle $ at $T=0$ is plotted in Fig. 3(a). The
oscillation is due to the alternate filling by electrons of the energy
levels with mainly spin-up and spin-down. A jump is visible at $\nu =12.6$
(correspondi ng to inverse magnetic field $0.162$T$^{-1}$) because of the
energy crossing.

The second part in G$_{s}$ arises from the first order in $H^{\prime }$, 
\begin{eqnarray}
G_{s}^{(1)} &=&\frac{e\eta _{R}}{8\pi \sqrt{2}}\sum_{n,s,n^{\prime
}=n+1,s^{\prime }}\frac{f_{ns}-f_{n^{\prime }s^{\prime }}}{\epsilon
_{ns}^{R}-\epsilon _{n^{\prime }s^{\prime }}^{R}}  \notag \\
&&\times \left( \sqrt{n}\sin 2\theta _{ns}\sin ^{2}\theta _{n^{\prime
}s^{\prime }}-\sqrt{n^{\prime }}\cos ^{2}\theta _{ns}\sin 2\theta
_{n^{\prime }s^{\prime }}\right)
\end{eqnarray}%
At $T=0$, if the two degenerate energy levels (crossing point in Fig. 2(a))
are partially occupied, $G_{s}^{z}$ may become divergent. Mathematically,
the resonance is given by the condition $2n<\nu <2n+1$ for the electron
filling factor $\nu $, with $n$ an integer satisfying the equation 
\begin{equation}
\sqrt{(1-g)^{2}+8n\eta _{R}^{2}}+\sqrt{(1-g)^{2}+8\left( n+1\right) \eta
_{R}^{2}}=2.  \label{condition}
\end{equation}%
From the above condition, for a system with any $\eta _{R}\neq 0,\,\eta
_{D}=0$, and $g_{s}>0$, there is a unique resonant magnetic field $B_{0}$
such that the resonant condition is satisfied. By symmetry, we obtain the
resonance condition for the system with a pure Dresselhaus coupling, which
is given by the solution for $n$ of the equation, 
\begin{equation}
\sqrt{(1+g)^{2}+8n\eta _{D}^{2}}+\sqrt{(1+g)^{2}+8\left( n+1\right) \eta
_{D}^{2}}=2.
\end{equation}%
Unlike the pure Rashba coupling case, there is no solution for any $g_{s}>0$
in the pure Dresselhaus coupling system. This is because the energy levels $%
\epsilon _{ns}^{D}$ and $\epsilon _{n^{\prime }s^{\prime }}^{D}$ with $%
n^{\prime }=n\pm 1$ do not cross over, so the pairs of the crossing levels
in Fig. 2(b) correspond to $n^{\prime }\neq n\pm 1$ and do not contribute to
the spin Hall conductance.

We have calculated the spin Hall conductance numerically. $G_{s}^{z}$ at $
T=0 $ is shown in Fig. 3(b). In addition to the oscillation in $1/B$ similar
to that of $\sigma _{z}$, there is a pronounced resonant peak at the filling 
$\nu =12.6$ . In Fig. 4, we show $G_{s}^{z}$ at several temperatures. The
height of the resonance peak increases drastically as the temperature
decreases below a few kelvin. In Fig. 5 , we show the $T$-dependence of the
height of the resonant peak and the two nearby side peaks. The
characteristic temperature for the occurrence of the peak can be estimated
to be the Zeeman energy $E_{Z}$, which is about $10K$ at the resonant field
for the parameters in Fig. 2. More explicit derivation of this will be given
in the next subsection.

\begin{figure}[tbp]
\includegraphics[width=8cm]{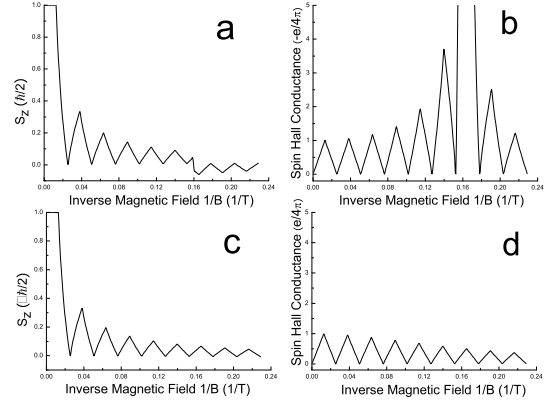}
\caption{{Average spin $S_{z}$ and spin Hall conductance as a function of
1/B at $T=0$. The parameters are the same as in Fig. 2. (a) and (b) for the
Rashba coupling, and (c) and (d) for the Dresselhaus coupling.}}
\end{figure}

\begin{figure}[tbp]
\includegraphics[width=8cm]{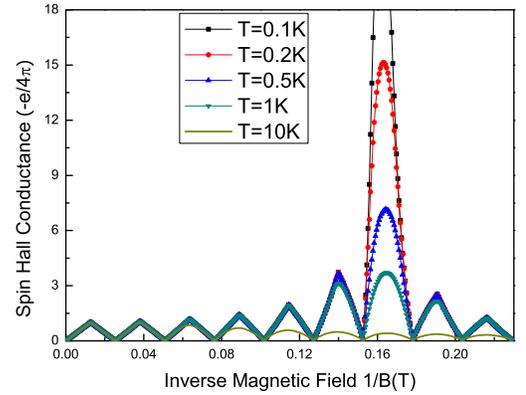}
\caption{{Spin Hall conductance v.s. $1/B$ at several temeratures for Rashba
coupling systems. The parameters are the same as those in Fig. 2.}}
\label{fig:fig.4}
\end{figure}

\begin{figure}[tbp]
\includegraphics[width=8cm]{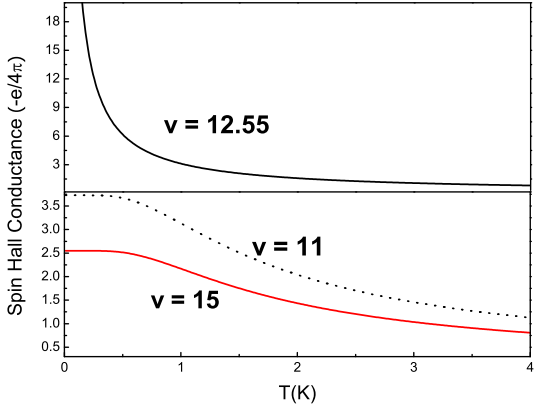}
\caption{{Temperature dependence of the height of the resonance peak and the
two side peaks in Fig. 4.}}
\end{figure}

\subsection{Non-ohmic spin Hall current and scaling behavior}

In this section we study the non-linear effect of the electric field to the
resonant spin Hall current and the scaling behavior. Since the resonance
originates from the interference of two degenerate levels near the Fermi
energy, we will focus on those two levels to examine the problem. As an
example, we shall consider In$_{0.53}$Ga$_{0.47}$As/In $_{0.52}$Ga$_{0.48}$
As with the parameters given in Fig. 2, in which case the resonance occurs
at the filling factor $\nu =12.6$ (see Fig. 3b) and the relevant two levels
are $\left\vert 1\right\rangle =\left\vert n=6,k,s=+1\right\rangle $ and $
\left\vert 2\right\rangle =\left\vert n+1=7,k,s=-1\right\rangle $. The
energy levels below the two levels are assumed to be fully filled, and all
levels above the two to be empty. This is valid if $\hbar \omega \gg $ $%
k_{B}T$ . The Hamiltonian is then, up to a constant, reduced to a $2\times 2$
matrix, 
\begin{equation}
H_{\text{reduced}}=\left( 
\begin{array}{cc}
\Delta \epsilon & v_{0} \\ 
v_{0} & -\Delta \epsilon%
\end{array}
\right)  \label{reduced}
\end{equation}
where 
\begin{eqnarray*}
\Delta \epsilon &=&(\epsilon _{6,+1}^{R}-\epsilon _{7,-1}^{R})/2, \\
v_{0} &=&\left\langle 2\right\vert H^{\prime }\left\vert 1\right\rangle
=-eEl_{b}\eta _{R}\cos \theta _{6,+1}\sin \theta _{7,-1}.
\end{eqnarray*}

\begin{figure}[tbp]
\includegraphics[width=8cm]{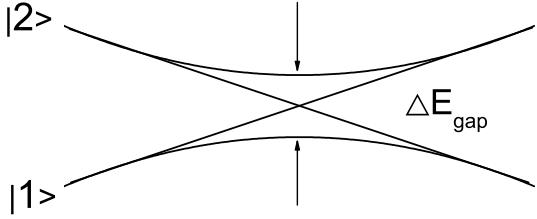}
\caption{{Schematic illustration of the energy shift due to the electric
field in the two degenerate levels near the resonant point. }}
\label{fig:fig.6}
\end{figure}

As we can see from the reduced Hamiltonian and from Fig. 6, the electric
field breaks the level degeneracy and opens an energy gap $\Delta
E_{gap}=2\left\vert v_{0}\right\vert $. Denoting the two eigenstates of the
reduced Hamiltonian by $\left\vert \Phi _{\pm }\right\rangle $, the spin
Hall current density is given by 
\begin{equation}
I_{s}=\frac{1}{2\pi l_{b}^{2}}\left( i_{-}f_{-}+i_{+}f_{+}\right)
\end{equation}%
where the Fermi-Dirac distribution $f_{\pm }=\{\exp \left( [\pm \sqrt{\left(
\Delta \epsilon \right) ^{2}+v_{0}^{2}}-\mu ]/k_{B}T\right) +1\}^{-1}$, with 
$f_{+}+f_{-}=\delta _{\nu }=\nu -2n$, $\mu $ the chemical potential, and $%
i_{\pm }=\left\langle \Phi _{\pm }\right\vert j_{x}^{z}\left\vert \Phi _{\pm
}\right\rangle $. The electric field and temperature dependences of the spin
current $I_{s}$ near the resonance point is plotted in Fig.7. At low
temperatures the resonant spin current approaches to a constant in a weak
electric field.

\begin{figure}[tbp]
\includegraphics[width=8.5cm]{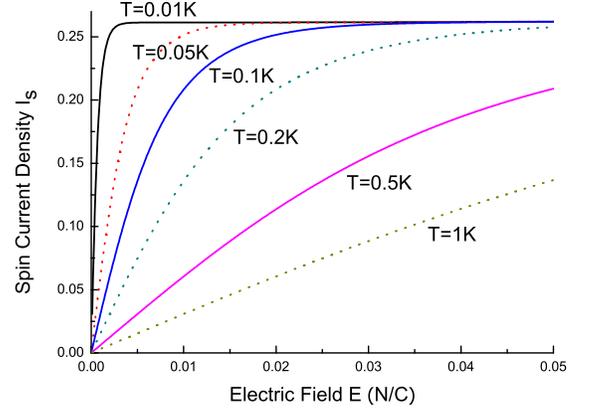}
\caption{{Resonant spin current density as a function of the electric field
at different temperatures. The spin current unit is $(-e/4\protect\pi %
)(N/C). $ The filling factor at the resonance is }$\protect\nu =12.6.$}
\end{figure}

Now we analyze the scaling behavior of the spin conductance near the
resonance point. For simplicity we limit our discussion to the case of $%
\delta _{\nu }<1$ and $g\ll 1$. Near the resonant point, $\Delta \epsilon
\approx -E_{Z}b$ where $E_{Z}=g\hbar \omega _{0}/2$ is the Zeeman energy and 
$b=(B-B_{0})/B_{0}$ is the reduced dimensionless magnetic field. Using the
identity 
\begin{equation}
f_{-}-f_{+}\equiv f_{-}\left( 1-f_{+}\right) \left[ 1-e^{-2\sqrt{\left(
\Delta \epsilon \right) ^{2}+v_{0}^{2}}/k_{B}T}\right] ,
\end{equation}%
we obtain the singular part of the spin Hall conductance to be 
\begin{equation}
G_{s}\simeq -\frac{\delta _{\nu }e}{4\pi }\frac{E_{Z}}{\sqrt{\left( \Delta
\epsilon \right) ^{2}+v_{0}^{2}}}\frac{f_{-}\left( 1-f_{+}\right) }{\left(
f_{-}+f_{+}\right) }\left[ 1-e^{-\frac{2\sqrt{\left( \Delta \epsilon \right)
^{2}+v_{0}^{2}}}{k_{B}T}}\right] .
\end{equation}%
where the factor $f_{-}\left( 1-f_{+}\right) /\left( f_{-}+f_{+}\right) $ is
a slowly varying function of $T$ ranging from 1 at low temperatures to $%
(1-\delta _{\nu }/2)/2$ at high temperatures. At low temperatures $G_{s}$ is
given by, 
\begin{equation}
G_{s}\simeq -\frac{e}{4\pi }\frac{\delta _{\nu }}{\left\vert b\right\vert }.
\end{equation}%
It is only a function of the reduced magnetic field and the excess part of
the filling factor from $2n$. At the resonant magnetic field, \textit{i.e.}, 
$b=0,$ the spin Hall current approaches with lowering temperature to a
constant, $I_{s}=-\frac{e}{4\pi }\delta _{\nu }E_{Z}/(el_{b}\eta _{R}\cos
\theta _{n,+1}\sin \theta _{n+1,-1})$ as can be seen in Fig. 7. Using the
resonance condition in Eq.(\ref{condition}), $B_{0}\approx 4nm^{2}c\beta
^{2}/g\hbar ^{3}$ ($n=6$) and using the fact that for large $n$, $n$ is
proportional to $1/B_{0},$ the resonant magnetic field $B_{0}\propto \beta /%
\sqrt{g}$ approximately. Thus the resonant spin current is proportional to 
\begin{equation}
I_{s}=-\frac{\delta _{\nu }e^{2}}{8\pi m^{2}c^{2}}\frac{gB_{0}^{2}}{\beta }%
\propto \delta _{\nu }\beta .
\end{equation}%
Therefore for a given filling factor, the larger the spin-orbit coupling $%
\beta $ is, the stronger the spin Hall resonance. The resulted spin Hall
conductance diverges at $T=0$ as 
\begin{equation}
G_{s}\simeq -\frac{\delta _{\nu }e}{4\pi }\frac{E_{Z}}{\left\vert
v_{0}\right\vert }=-\frac{\delta _{\nu }e^{2}}{8\pi m^{2}c^{2}}\frac{%
gB_{0}^{2}}{\beta }\frac{1}{E}.
\end{equation}%
At temperatures $k_{B}T>\sqrt{\left( \Delta \epsilon \right) ^{2}+v_{0}^{2}}$
, 
\begin{equation}
G_{s}\simeq -\frac{\delta _{\nu }(1-\delta _{\nu }/2)e}{4\pi }\frac{2E_{Z}}{%
k_{B}T}.
\end{equation}%
and the integral 
\begin{equation*}
\int G_{s}db\rightarrow -\frac{\delta _{\nu }e}{2\pi }\left( \ln \frac{2E_{Z}%
}{k_{B}T}\right) ,
\end{equation*}%
This integral reflects the weight of the resonant peak of the spin Hall
conductance.

Since the method used in this subsection is beyond perturbation theory, we
conclude that the resonance spin Hall conductance we predict is not an
artifact of the perturbation method. Instead, the resonance is caused by the
interference between the two degenerate energy levels at the Fermi energy.

\section{Systems with both Rashba and Dresselhaus couplings}

In this section we briefly discuss the resonance in the spin Hall
conductance in systems with both Rashba and Dresselhaus couplings. The
Hamiltonian including the electric potential reads, 
\begin{equation}
H=H_{0}(E)+H^{\prime }
\end{equation}
with $H^{\prime }=-eE\,l_{b}(\eta _{D}\sigma _{x}+\eta _{R}\sigma _{y})$. In
this case $H_{0}$ is not solvable analytically. A state $\left\vert
n_{0},\downarrow \right\rangle $ (in the basis of the Landau levels with $
\eta _{D}=\eta _{R}=0$) is coupled to $\left\vert n_{0}+1,\uparrow
\right\rangle $ via the Rashba coupling, which is further coupled to $
\left\vert n_{0}+2,\downarrow \right\rangle $ due to the Dresselhaus
coupling. In this way, a Landau level is coupled to infinite number of other
Landau levels, and the analytic solution is not available. The problem,
however, may be approximately solved by using perturbation theory to treat $%
\eta _{R}$ and $\eta _{D}$ as small parameters. This is equivalent to the
limit $B\rightarrow \infty $, since $\eta _{D,R}\propto 1/\sqrt{B}$. For the
sample we consider in the present paper given in Fig. 2, $\eta
_{R}^{2}=0.004\ll 1$ at the resonant field $B\approx 6.1$ tesla. In the
absence of the electric field, the single particle energy, up to the second
order in $\eta _{R}$ and $\eta _{D}$, is given by

\begin{subequations}
\label{two-e}
\begin{eqnarray}
\frac{\epsilon _{n_{0}\uparrow }}{\hbar \omega } &=&n_{0}+\frac{1-g}{2}+ 
\frac{2n_{0}\eta _{R}^{2}}{1-g}-\frac{2(n_{0}+1)\eta _{D}^{2}}{1+g} \\
\frac{\epsilon _{n_{0}\downarrow }}{\hbar \omega } &=&n_{0}+\frac{1+g}{2}+ 
\frac{2n_{0}\eta _{D}^{2}}{1+g}-\frac{2(n_{0}+1)\eta _{R}^{2}}{1-g}.
\end{eqnarray}
Note that the mixed term of $\eta _{R}\eta _{D}$ does not appear in the
perturbation to the second order. The two levels become degenerate if the
following equation is satisfied, 
\end{subequations}
\begin{equation}
\frac{g}{2(2n_{0}+1)}=\frac{\eta _{R}^{2}}{1-g}-\frac{\eta _{D}^{2}}{1+g}.
\label{perturbation}
\end{equation}
It follows that a necessary condition for the resonant spin Hall current is $
\eta _{R}^{2}/\eta _{D}^{2}>(1-g)/(1+g)\approx 1$, for $g\ll 1$. At $\eta
_{D}=0$ and in the limit $\eta _{R}\ll 1$, Eq. (\ref{perturbation}) is
consistent with Eq. (\ref{condition}) for the resonant condition we derived
for the pure Rashba system. Alternatively the resonant magnetic field is 
\begin{equation}
B_{0}\approx \frac{2(2n_{0}+1)}{g}\frac{m^{2}c}{e\hbar ^{3}}\left( \beta
^{2}-\alpha ^{2}\right) .
\end{equation}
The large number $n_{0}$ increases with $1/B_{0}$ for a specific density of
particles. Thus for a certain Rashba coupling the increasing of Dresselhaus
coupling will decrease the resonant magnetic field $B_{0}.$ The singular
part of the spin Hall conductance can be studied by examining the two level
system in the presence of an electric field as we described in Section IIIC.
At the resonant point and at low temperature,

\begin{equation}
G_{s}=-\frac{\delta _{\nu }e^{2}\hbar ^{2}}{8\pi m^{2}c^{2}}\frac{gB_{0}^{2} 
}{\sqrt{\alpha ^{2}+\beta ^{2}}}\frac{1}{E}.
\end{equation}
As the Dresselhaus coupling increases from zero, the resonance is shifted to
lower magnetic fields, the resonance occurs at higher Landau levels with a
weaker resonant strength.

\section{Summary and Discussions}

In summary, we have studied the spin Hall effect in a two-dimensional
electron system with spin-orbit couplings in a strong perpendicular magnetic
field. In systems with the Rashba coupling dominating over the Dresselhaus
coupling, there is a resonant magnetic field at which the spin Hall
conductance diverges at low temperature and low electric field. The physics
for this resonance is the energy level crossing of the two Landau levels due
to the competition of the Zeeman splitting and the Rashba coupling. For a
given system, there is a unique resonant magnetic field, at which the two
Landau levels become degenerate at the Fermi energy. In this case, some
physical properties may show singularity. As studied earlier, the spin
polarization will change its sign as the magnetic field is varied passing
through the resonant field. Namely the magnetic susceptibility is divergent.
The spin Hall conductance is another singular response due to this level
crossing. When an infinitesimally weak d.c. electric field is applied in the
plane, the two degenerate Landau levels are split accordingly and a finite
spin Hall current is induced. The resonance is macroscopic in the sense that
a huge number of the states in the same Landau level are involved in the
process. We have calculated the temperature and electric field dependences
of the resonance. The characteristic temperature for the resonant spin Hall
current is of order of the Zeeman energy. As the temperature decreases, the
height of the resonance peak diverge like $\propto 1/T$ and the weight
diverges like $\propto \ln T$. While the spin orbit coupling has a dramatic
effect on the spin Hall conductance, the charge Hall conductance is not
affected and remains quantized. The spin Hall current is non-linear with the
electric field at the resonant field. At low temperatures, the spin Hall
current rapidly rises linearly with the electric field and saturates at
higher electric fields. At $T=0$, the spin Hall conductance diverges as $1/E$
at resonance. Near the resonant magnetic field $B_{0}$, it is $\propto
1/\left\vert B-B_{0}\right\vert $. Contrary to the Rashba coupling, the
Dresselhaus coupling further increases the Zeeman energy splitting to
suppress the effect of the Rashba coupling. The strength of the Rashba
coupling necessary to surpass the Dresselhaus coupling by in order to have
the resonant spin Hall current was estimated by using a perturbation method
treating the couplings as small parameters. This is accurate as long as the
Zeeman energy is much smaller than then cyclotron frequency.

We have assumed no potential disorder in our theory. The effects of disorder
in 2DEG with Rashba coupling, especially in a strong magnetic field, is not
well understood at this point.\cite{Xiong} Nevertheless, it seems reasonable
to assume that the spin-orbit coupling does not change the effects of
disorder qualitatively. This is likely to be the case in the presence of a
strong magnetic field, which ensures extended states in the Landau levels
when the disorder is not sufficiently strong as evidenced by the
experimentally observed quantization of the Hall conductance. We then assume
that the disorder gives rise to broadening of the Landau level and
localization so that the extended states in a Landau levels are separate in
energy from those in the next one by localized states. Inspection of the
spin-orbit coupling shows that Laughlin's gauge argument still holds\cite%
{laughlin,halperin}, and each Landau level with its extended states
completely filled contribute $e^{2}/h$ to the charge Hall conductance. Thus
we conclude that the quantum Hall conductance remains intact with the
spin-orbit interaction, except at the special degeneracy point. As the Fermi
energy varies across this degenerate extended state, the charge Hall
conductance $G_{c}$ is expected to change by $2e^{2}/h$, instead of $e^{2}/h$
for the other extended levels. This fact can be used experimentally to
determine the Rashba interaction induced degeneracy discussed in this paper.

\begin{acknowledgments}
This work was in part supported by the Research Grant Council in Hong Kong
(SQS), NSF ITR Grant No. 0223574 (FCZ), and DOE/DE-FG02-04ER46124 (XCX).
\end{acknowledgments}


\begin{thebibliography}{99}
\bibitem{Prinz98Science} G. A. Prinz, Science \textbf{282}, 1660 (1998)

\bibitem{Wolf01Science} S. A. Wolf, D. D. Awschalom, R. A. Buhrman, J. M.
Daughton, S. von Molnar, M. L. Roukes, A. Y. Chtchelkanova, and D. M.
Treger, Science \textbf{294}, 1488 (2001).

\bibitem{Awschalom02} D. Awschalom, D. Loss, and N. Samarth (.eds),
Semiconductor Spintronics and Quantum Computation (Springer, Berlin, 2002).

\bibitem{Murakami03Science} S. Murakami, N. Nagaosa, and S. C. Zhang,
Science \textbf{301}, 1348 (2003); Phys. Rev. B \textbf{69}, 235206 (2004).

\bibitem{Sinova04} J. Sinova, D. Culcer, Q. Niu, N. A. Sinitsyn, T.
Jungwirth, and A. H. MacDonald, Phys. Rev. Lett. \textbf{92}, 126603 (2004)

\bibitem{Shen04prl} S. Q. Shen, M. Ma, X. C. Xie, and F. C. Zhang, Phys.
Rev. Lett. \textbf{92}, 256603 (2004).

\bibitem{Shen03xxx} S. Q. Shen, Phys. Rev. B \textbf{70}, 081311(R) (2004).

\bibitem{Sinitsyn03xxx} N. A. Sinitsyn, E. M. Hankiewich, W. Teizer, and J.
Sinova, Phys. Rev. B \textbf{70}, 081312(R) (2004).

\bibitem{Culcer03xxx} D. Culcer, J. Sinova, N. A. Sinitsyn, A. H. MacDonald,
and Q. Niu, Phys. Rev. Lett. \textbf{93}, 046602 (2004).

\bibitem{Schliemann04prb} J. Schliemann and D. Loss, Phys. Rev. B \textbf{69}
, 165315 (2004).

\bibitem{Hu03xxx} J. Hu, B.A. Bernevig, and C. Wu, Int. J. Mod. Phys. B 
\textbf{17}, 5991 (2003).

\bibitem{Rashba03prb} E. I. Rashba, Phys. Rev. B \textbf{68}, 241315(R)
(2003).

\bibitem{Hulb04xxx} L. Hu, J. Gao, and S. Q. Shen, cond-mat/0401231; X. Ma,
L. Hu, R. Tao and S. Q. Shen, cond-mat/0407419.

\bibitem{Dyakonov71} M. I. Dyakonov and V. I. Perel, Phys. Lett. A \textbf{\
35 }, 459 (1971).

\bibitem{Hirsch99} J. E. Hirsch, Phys. Rev. Lett. \textbf{83}, 1834 (1999).

\bibitem{Zhang00prl} S. Zhang, Phys. Rev. Lett. \textbf{85}, 393 (2000)

\bibitem{Hu03prb} L. Hu, J. Gao, and S. Q. Shen, Phys. Rev. B \textbf{68},
115302 (2003); \textbf{68}, 153303 (2003).

\bibitem{Shen97pla} For example, see S. Q. Shen, Phys. Lett. A \textbf{235},
403 (1997); S. Q. Shen and X. C. Xie, Phys. Rev. B \textbf{67}, 144423
(2003).

\bibitem{Klitzing80} K. v. Klitzing, G. Dorda, and M. Pepper, Phys. Rev.
Lett. \textbf{45}, 494 (1980).

\bibitem{Tsui82} D. C. Tsui, H. L. Stormer, and A. C. Gossard, Phys. Rev.
Lett. \textbf{48}, 1559 (1982).

\bibitem{Prange87} R. E. Prange and S. M. Girvin (.eds), The Quantum Hall
Effect, (Springer, Berlin, 1987); S. Das Sarma and A. Pinczuk (.eds),
Perspectives in Quantum Hall Effects, (Wiley, 1997).

\bibitem{Nitta97prl} J. Nitta, T. Akazaki, H. Takayanagi, and T. Enoki,
Phys. Rev. Lett. \textbf{78}, 1335 (1997).

\bibitem{Heida98prb} J. P. Heida, B. J. van Wees, J. J. Kuipers, T. M.
Klapwijk, and G. Borghs, Phys. Rev. B \textbf{57}, 11911 (1998).

\bibitem{Dresselhaus55} G. Dresselhaus, Phys. Rev. \textbf{100}, 580 (1955).

\bibitem{Rashba60} E. I. Rashba, Fiz. Tverd. Tela (Leningrad) \textbf{2},
1224 (1960) [Sov. Phys. Solid State \textbf{2}, 1109 (1960)]; Y. A. Bychkov
and E. I. Rashba, J. Phys. C \textbf{17}, 6039 (1984).

\bibitem{Grundler00} D. Grundler, Phys. Rev. Lett. \textbf{84}, 6074 (2000).

\bibitem{Jusserand95} B. Jusserand, D. Richards, G. Allan, C. Priester, and
B. Etienne, Phys. Rev. B \textbf{51}, 4707 (1995).

\bibitem{Knap96} W. Knap, C. Skierbiszewski, A. Zduniak, E.
Litwin--Staszewska, D. Bertho, F. Kobbi, J. L. Robert, G. E. Pikus, F. G.
Pikus, S. V. Iordanskii, V. Mosser, K. Zekentes, Yu. B. Lyanda--Geller,
Phys. Rev. B \textbf{53}, 3912 (1996).

\bibitem{Miller03} J. B. Miller, D. M. Zumb\"{u}hl, C. M. Marcus, Y. B.
Lyanda-Geller, D. Goldhaber-Gordon, K. Campman, and A. C. Gossard, Phys.
Rev. Lett. \textbf{90}, 076807 (2003).

\bibitem{Ganichev04prl} S. D. Ganichev, V. V. Bel'kov, L. E. Colub, E. L.
Ivchenko, Petera Schneider, S. Giglberger, J. De Boeck, G. Borghs, W.
Wegscheider, D. Weiss, and W. Prettl, Phys. Rev. Lett. \textbf{92}, 256601
(2004).

\bibitem{note1} A hole in semiconductor carries a positive charge $e$.

\bibitem{Das90} B. Das, S. Datta, and R. Reifenberger, Phys. Rev. B \textbf{%
\ \ 41}, 8278 (1990).

\bibitem{Falko93} V. I. Falko, Phys. Rev. Lett. \textbf{71}, 141 (1993).

\bibitem{Schliemann03prb} J. Schliemann, J. C. Egues, and D. Loss, Phys.
Rev. B \textbf{67}, 085302 (2003).

\bibitem{Xiong} Y. Xiong and X. C. Xie, cond-mat/0403083; J. Inoue, G. E. W.
Bauer, and L. W. Molenkamp, Phys. Rev. B 70, 041303 (2004); S. Murakami,
Phys. Rev. B 69, 241202 (2004); E. G. Mishchenko, A. V. Shytov, and B. I.
Halperin, cond-mat/0406730; K. Nomura, J. Sinova, T. Jungwirth, Q. Niu, and
A. H. MacDonald, cond-mat/0407279; L. Sheng, D. N. Sheng, and C. S. Ting,
cond-mat/0409038.

\bibitem{laughlin} R. B. Laughlin, Phys. Rev. B \textbf{23}, 5632 (1981).

\bibitem{halperin} B. I. Halperin, Phys. Rev. B \textbf{25}, 2185 (1982).
\end{thebibliography}
\end{document}